\documentclass[twocolumn,aps,prb,reprint,superscriptaddress]{revtex4}

\usepackage{graphicx}
\usepackage{bm}
\begin{document}
\newcommand{\Eqn}[1]{&\hspace{-0.5em}#1\hspace{-0.5em}&}
\newcommand{\simg}{\stackrel{>}{_\sim}}
\newcommand{\siml}{\stackrel{<}{_\sim}}


\title{Cooperative effects of Coulomb and electron-phonon interactions in the two-dimensional 16-band
$d$-$p$ model for iron-based superconductors}


\author{Yuki Yanagi}
\affiliation{Department of Physics, Niigata University, Ikarashi,
Niigata 950-2181, Japan} 
\author{Youichi Yamakawa}
\affiliation{Department of Physics, Niigata University, Ikarashi,
Niigata 950-2181, Japan} 
\author{Naoko Adachi}
\affiliation{Department of Physics, Niigata University, Ikarashi,
Niigata 950-2181, Japan} 
\author{Yoshiaki \=Ono}
\affiliation{Department of Physics, Niigata University, Ikarashi,
Niigata 950-2181, Japan} 


\date{\today}

\begin{abstract}
We study  the
 electronic states and the superconductivity in the
 two-dimensional 16-band $d$-$p$  model coupled with $A_{1g}$, $B_{1g}$
 and 
 $E_{g}$ local phonons and obtain the
  rich phase diagram including the magnetic, charge and orbital
 ordered phases on the parameter plane of the Coulomb and electron-phonon
 interactions. When the electron-phonon interaction is dominant, the
 charge fluctuations induce the $s_{++}$-wave superconductivity, while
 when the Coulomb interaction is dominant, the magnetic fluctuations
 induce the $s_{\pm}$-wave superconductivity. 
 Remarkably,   the orbital fluctuations are enhanced due to the
 cooperative effects of the Coulomb and electron-phonon 
 interactions and induce the
 $s_{++}$-wave and the nodal $s_{\pm}$-wave superconductivities. 
\end{abstract}

\pacs{74.20.Rp, 74.25.Dw, 74.70.Xa, 74.20.Mn}

\maketitle

\section{Introduction}

The recently discovered iron-based superconductors\cite{kamihara_1,kamihara}
RFe$Pn$O$_{1-x}$F$_x$ (R=Rare Earth, $Pn$=As, P) with a transition
temperature $T_c$ exceeding 50K\cite{chen_1,ren_1,ren_2,chen_2,ren_3} 
have attracted much
attention. 
At present, there are following four families of the iron-based
superconductors: RFeAsO with ZrCuSiAs-type structure (1111 system),
BaFe$_2$As$_2$ with ThCr$_2$Si$_2$-type structure (122 system)\cite{rotter,rotter_2},
 LiFeAs and NaFeAs
with PbFCl-type structure (111 system)\cite{x_c_wang,pitcher,tapp} and Fe(Se,Te) (11
system)\cite{hsu,yeh}.  These systems have similar 
conducting Fe-pnictogen (-chalcogen) planes and the resulting electronic structures
predicted by the first principle calculations are similar to those for each
families\cite{lebegue,singh,haule,xu,boeri,nekrasov,nekrasov_2,subedi,ma}. The energy bands near the Fermi level mainly constructed
by the Fe-$3d$ orbitals are heavily entangled and there are two or
three concentric hole Fermi surfaces (FSs) around the $\Gamma$-point
$[\bm{k}=(0,0)]$ and the elliptical electron FSs around the $M$-point
$[\bm{k}=(\pi,\pi)]$. These features are observed by the angle
resolved photoemission spectroscopy (ARPES) in several compounds\cite{ding,lu,yi}. 
 Despite the similarities of the electronic
structures for the four families, it seems that the details of the gap
structures are different from system to 
system as mentioned below and the pairing state together with the mechanism of 
the superconductivity for the iron-based superconductors is still controversial.

As for the 1111 system, the F nondoped compound LaFeAsO exhibits the structural transition from
tetragonal (P4/nmm) to orthorhombic (Cmma) phase 
at a transition temperature $T=$155K and stripe-type antiferromagnetic (AFM)
order 
at $T=134\mathrm{K}$ with a magnetic moment
$\sim0.36\mu_{B}$\cite{cruz} at low temperature. With increasing F doping, the system becomes metallic and the
AFM order disappears\cite{kamihara}, and then, the
superconductivity emerges for $x\sim0.1$ with $T_c\sim26\mathrm{K}$.
 Rare-earth substitution compounds exhibit superconducting transition
 with higher $T_c$\cite{chen_1,ren_1,ren_2,chen_2,ren_3}. 
The NMR Knight shift measurements
revealed that the superconductivity of the systems is the spin-singlet
pairing\cite{matano,kawabata_1}. Fully-gapped superconducting states have been predicted by various experiments
such as the penetration depth\cite{hashimoto}, the specific
heat\cite{mu},  and the impurity effect on 
$T_c$\cite{kawabata_1,karkin}.  
 In contrast to the above mentioned experiments, the NMR relaxation rate
 shows the power low behavior $1/T{_1}\propto T^3$ below
 $T_{c}$\cite{nakai}, suggesting the nodal or  highly anisotropic gap
 structure. The other NMR measurements\cite{kobayashi}, however, revealed $1/T{_1}\propto T^6$ below
 $T_c$ and there is still controversy. 

 The parent compound of the 122 system BaFe$_2$As$_2$ shows  the structural transition from
tetragonal (I4/mmm) to orthorhombic (Fmmm) phase 
and the stripe-type AFM
order simultaneously at a transition temperature
$T=$140K\cite{rotter_2,huang}, where the magnetic moment is 
about $0.87\mu_{B}$ at low temperature\cite{huang}.
Both the electron and hole doping by the substitution Co for Fe and K
for Ba 
 induce the superconductivity\cite{rotter,sefat}. The 
$T^5$-dependence of $1/T_1$\cite{yashima}, the exponential behavior of the
penetration depth\cite{hashimoto_3} and  the ARPES\cite{liu,ding,kondo}
suggest the fully-gapped superconductivity. 
The chemical pressure by substituting P for As in BaFe$_2$(As$_{1-x}$P$_x$)$_2$ also leads to the superconductivity with $T_c$ up to 30K\cite{jiang}, where
 the specific heat, the penetration
depth, the thermal conductivity and NMR $1/T_1$ imply the nodal or highly
anisotropic gap structures.\cite{hashimoto_2,nakai_2}.

 In 11 system, FeTe shows
 the another type of the AFM order with ordering vector $\bm{q}=(\pi,0)$\cite{fang,bao},  where the
 magnetic moment is about 2.03$\mu_B$\cite{fang,bao}.
 On the other hand, FeSe does not exhibit the magnetic order but
 superconducting transition at $T\sim8$K\cite{hsu}.  The thermal
 conductivity\cite{dong} 
  suggests the fully-gapped superconducting state.

Theoretically, Mazin \textit{et al.} suggested
that the fully-gapped $s$-wave pairing whose order parameter
changes its sign between the hole FSs and the electron FSs
($s_{\pm}$-wave pairing) is
favored due to the stripe-type AFM spin fluctuations\cite{mazin}.
According to the
weak coupling approaches based on multi-orbital Hubbard
models\cite{kuroki_1,kuroki_2,kuroki_3,nomura_2,ikeda,ikeda_2,wang_1,graser,maier,yao,stanescu,cvetkovic},
and those based on the $d$-$p$ model\cite{yamakawa_1,yamakawa_2,yamakawa_4},
the $s_{\pm}$-wave pairing seems to be the promising candidate for the pairing state
in the iron-based superconductors.  It is shown that the $s_{\pm}$-wave pairing is realized also in the
strong coupling region by the mean field study based on 
the $t$-$J_1$-$J_2$ model\cite{seo} and the exact diagonalization study
based on the one-dimensional two-band
Hubbard model\cite{sano}. The $s_{\pm}$-wave state mediated by the spin
fluctuations seems to be consistent with
many experiments. However, the theoretical analysis of the nonmagnetic impurity
effects based on the 5-band Hubbard model shows that the
$s_{\pm}$-wave state is very fragile against nonmagnetic
impurities\cite{onari}. This is in
contradiction to the experimental results that the superconductivity for
the iron-based superconductors is
robust against nonmagnetic impurities. Therefore, the fully-gapped
$s$-wave state without sign reversing ($s_{++}$-wave
state) is considered to be another promising candidate for the pairing
state in the 
iron-based superconductors. 

 In the previous papers\cite{yamakawa_1,yamakawa_2,yamakawa_4,yamakawa_5}, we have investigated
the electronic states of the Fe$_2$As$_2$ plane in iron-based
superconductors on the basis of the two-dimensional
16-band $d$-$p$ model which includes the
Coulomb interaction on a Fe site: the intra- and inter-orbital direct
terms $U$ and $U'$, the Hund's coupling $J$ and the pair-transfer
$J'$. Using the random phase
approximation (RPA), we have found that, for $U>U'$,
the $s_{\pm}$-wave superconductivity is realized due to  the spin
fluctuations with $\bm{q}\sim(\pi,\pi)$, while for $U<U'$, the $s_{++}$-wave state
is realized due to the orbital fluctuations with $\bm{q}=(0,0)$\cite{yamakawa_5}. In
addition, we suggest that the electron-phonon interaction enhances the orbital
fluctuations and plays the significant role in the realization of the
$s_{++}$-wave superconductivity in the realistic parameter region $U>U'$. 
 In the recent Raman spectroscopy, it is shown that the
electron-phonon coupling constant for $A_{1g}$ and $B_{1g}$ modes are
larger ($\lambda_{A_{1g}}, \lambda_{B_{1g}}\sim 0.5$)\cite{rahlenbeck} than those
predicted by the first principle calculations ($\lambda\sim 0.21$)\cite{boeri}.
Then, it is important to investigate the effects of the electron-phonon
interaction on the electronic states and the superconductivity based on
the microscopic model.

Recently,  Kontani and Onari have investigated the 5-band
Hubbard-Holstein model which includes the Coulomb interaction and the
electron-phonon interaction due to the $B_{1g}$ and $E_g$ phonons at the
zone center by using the RPA and have shown that the $E_g$ phonons
drastically enhance the orbital
fluctuations and the $s_{++}$-wave superconductivity is realized by the
orbital fluctuations for the realistic values of the electron-phonon coupling\cite{kontani}.
In ref. 68, they derive the electron-phonon
coupling 
by calculating the electrostatic potential variance for Fe-$3d$ electrons
from the four surrounding As$^{3-}$ ions due to the oscillations of the Fe
atoms assuming that the spatial extensions of the Fe-$3d$ like Wannier
functions are small. The spatial
extensions of the Fe-$3d$ like Wannier orbitals in the 5-band model,
however, are very large, e. g., $\langle r^2 \rangle$-$\langle r\rangle^2\sim 5.37$\AA$^2$ for
 $d_{x^2-y^2}$ orbital in LaFeAsO, \cite{vildosola,nakamura,miyake} in
 contrast to their assumption. 
  On the other hand, in the effective model
which includes both the Fe $3d$ orbitals
and the As $4p$ orbitals, so called $d$-$p$ model, the spatial
extensions of the Wannier functions are considered to be largely reduced\cite{vildosola,nakamura,miyake}. 
 Therefore,
 theoretical studies on the electron-phonon interaction based on the $d$-$p$ model, are 
 highly desired.  

In the present
paper, we investigate the effects of the electron-phonon interaction on
the electronic states and superconductivity  based on the two-dimensional 
16-band $d$-$p$  model, where the $A_{1g}$, $B_{1g}$ and $E_{g}$ phonons at the zone
center are considered.
Solving the linearized Eliashberg equation with the pairing 
interaction obtained by using the RPA, we obtain the phase diagram on
the parameter plane of the Coulomb and electron-phonon interactions.

The paper is organized as follows. In Sec. II, we introduce the the two-dimensional 
16-band $d$-$p$  model coupled with the $A_{1g}$, $B_{1g}$ and $E_{g}$
 local phonons and explain the formulation of the RPA with the Coulomb
 and electron-phonon interactions in the multi-orbital system. In
 Sec. III, we show the numerical results of the charge-orbital
 susceptibility and the gap function for the various values of the
 Coulomb and electron-phonon interactions. The linearized Eliashberg
 equation is solved and we obtain the phase diagram.  Finally, we summarize the paper in Sec IV.

\section{Model and Formulation}
Our model Hamiltonian is 
 the two-dimensional 16-band $d$-$p$  model\cite{yamakawa_1,yamakawa_2,yamakawa_3,yamakawa_4,yamakawa_5} coupled with local phonons, 
where $3d$ orbitals ($d_{3z^2-r^2}$, $d_{x^2-y^2}$, $d_{xy}$, $d_{yz}$, $d_{zx}$) of two Fe
atoms (Fe$^1$=$A$, Fe$^2$=$B$) and $4p$ orbitals ($p_{x}$, $p_{y}$, $p_{z}$) of two As atoms are
explicitly included. It is noted that
$x, y$ axes are directed along second nearest Fe-Fe
bonds. Here, we number the Fe-$3d$ orbitals as follows:
$d_{3z^2-r^2}$(1), $d_{x^2-y^2}$(2), $d_{xy}$(3), $d_{yz}$(4), 
 $d_{zx}$(5). 

\begin{figure}[t]
\begin{center}
\includegraphics[width=60mm]{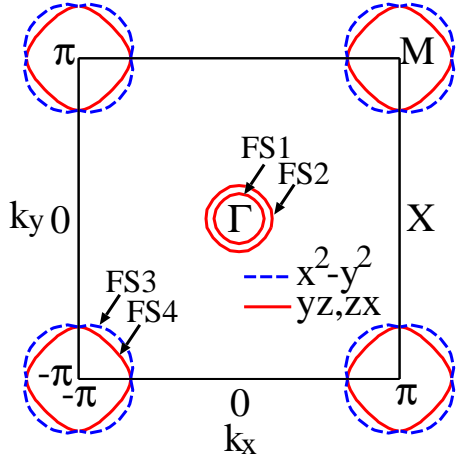}
\end{center}
\caption{(Color online) FSs obtained from the $d$-$p$ model for
 $x=0.1$. The solid and dashed lines show the FSs which have
 mainly $d_{yz}, d_{zx}$ and $d_{x^2-y^2}$ orbital character, respectively.
\label{FS}}
\end{figure}

The total Hamiltonian of the $d$-$p$ model is given by
\begin{equation}
H=H_0+H_\mathrm{int}+H_{ph}+H_{el-ph}, \label{eq_H}
\end{equation}
where $H_0$,  $H_\mathrm{int}$, $H_{ph}$ and $H_{el-ph}$ are the
kinetic, Coulomb interaction, phonon and electron-phonon interaction
parts of the Hamiltonian, respectively.
The kinetic part of the Hamiltonian is given by the following tight-binding Hamiltonian,
\begin{eqnarray}
H_0&=&\sum_{i,\ell,\sigma}\hspace{-1mm}
\varepsilon^d_{\ell}d^{\dag}_{i\ell\sigma}d_{i\ell\sigma}
+\hspace{-1mm}\sum_{i,m,\sigma}\hspace{-1mm}\varepsilon^p_{m}p^{\dag}_{im\sigma}p_{im\sigma} \nonumber \\ 
&+&\sum_{i,j,\ell,\ell',\sigma}\hspace{-1mm}t^{dd}_{i,j,\ell,\ell'}d^{\dag}_{i\ell\sigma}d_{j\ell'\sigma}
 +\hspace{-1mm} \sum_{i,j,m,m',\sigma}\hspace{-1mm}t^{pp}_{i,j,m,m'}p^{\dag}_{im\sigma}p_{jm'\sigma} \nonumber\\
&+&\sum_{i,j,\ell,m,\sigma}\hspace{-1mm}t^{dp}_{i,j,\ell,m}d^{\dag}_{i\ell\sigma}p_{jm\sigma}+h.c. \label{d-p}, 
\end{eqnarray}
where $d_{i\ell\sigma}$ is the annihilation operator for Fe-$3d$ electrons with spin
$\sigma$ in the orbital $\ell$ at the site $i$ and $p_{im\sigma}$ is the annihilation
operator for As-$4p$ electrons with spin
$\sigma$ in the orbital $m$ at the site $i$. In eq. (\ref{d-p}), the
transfer integrals $t^{dd}_{i,j,\ell,\ell'}$, $t^{pp}_{i,j,m,m'}$,
$t^{dp}_{i,j,\ell,m}$ and the atomic energies $\varepsilon^d_{\ell}$,
$\varepsilon^p_{m}$ are determined so as to fit both the energy and the
weights 
of orbitals for each band obtained from the
tight-binding approximation to
those from the density functional calculation for LaFeAsO and are listed in
refs. 64 and 70.
 The doping concentration $x$
corresponds to the number of electrons per unit cell $n=24+2x$ in the present
model. The FSs for $x=0.1$ are shown in Fig. \ref{FS} and we see the two
hole FSs (FS1 and FS2) and the two electron FSs (FS3 and FS4) as
predicted by the first principle calculations\cite{lebegue,singh,haule,xu,boeri}.

The Coulomb interaction part of the Hamiltonian is given as follows, 
\begin{eqnarray}
H_{\rm int}&=&\frac{1}{2}U\sum_{i}\sum_{\ell}\sum_{\sigma\neq\bar{\sigma}}
d^{\dag}_{i\ell\sigma}d^{\dag}_{i\ell\bar{\sigma}}
d_{i\ell\bar{\sigma}}d_{i\ell\sigma} \nonumber \\
&+&\frac{1}{2}U'\sum_{i}\sum_{\ell\neq\bar{\ell}}\sum_{\sigma,\sigma'}
d^{\dag}_{i\ell\sigma}d^{\dag}_{i\bar{\ell}\sigma'}
d_{i\bar{\ell}\sigma'}d_{i\ell\sigma} \nonumber \\
&+&\frac{1}{2}J\sum_{i}\sum_{\ell\neq\bar{\ell}}\sum_{\sigma,\sigma'}
d^{\dag}_{i\ell\sigma}d^{\dag}_{i\bar{\ell}\sigma'}
d_{i\ell\sigma'}d_{i\bar{\ell}\sigma} \nonumber \\
&+&\frac{1}{2}J'\sum_{i}\sum_{\ell\neq\bar{\ell}}\sum_{\sigma\neq\bar{\sigma}}
d^{\dag}_{i\ell\sigma}d^{\dag}_{i\ell\bar{\sigma}}
d_{i\bar{\ell}\bar{\sigma}}d_{i\bar{\ell}\sigma}, \label{eq_H_int}
\end{eqnarray}
where $U$ and $U'$ are the intra- and inter-orbital direct terms,
respectively, and $J$ and $J'$ are the Hund's coupling and the
pair-transfer, respectively. We assume that the relations between Coulomb
matrix elements $U=U'+2J$ and $J=J'$ are satisfied throughout the present paper.  

\begin{figure}[t]
\begin{center}
\includegraphics[width=27mm,angle=-90]{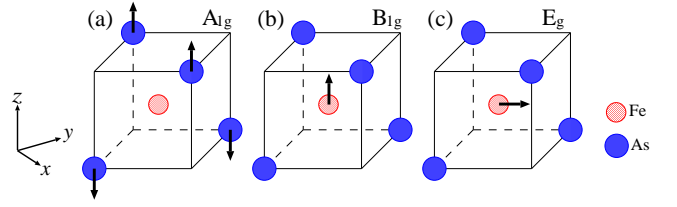}
\caption{(Color online) Schematic figure of the oscillations of Fe and As
 atoms in the (a) $A_{1g}$, (b) $B_{1g}$ and, (c) $E_g$ modes,
 respectively. The small and large spheres denote Fe and As atoms, respectively.
\label{fig_phonon}}
\end{center}
\end{figure}

Now we consider the effect of the phonon and the electron-phonon
interaction parts of the Hamiltonian $H_{ph}$ and $H_{el-ph}$. By performing the group theoretical
analysis, it is found that there are 
14 kinds of the optical phonon modes at the zone center. In the present paper,
we consider the $A_{1g}$,
$B_{1g}$ and $E_{g}$ phonon modes in which As atoms oscillate along the
$z$-axis,  Fe atoms oscillate along the $z$-axis and Fe atoms oscillate in
the $x$-$y$ plane, respectively (see Fig. \ref{fig_phonon}).  Here and
hereafter, we neglect the momentum dependence of
the electron-phonon couplings and that of the phonon frequencies for simplicity. The resulting
phonon and the electron-phonon 
interaction parts of the Hamiltonian are given as,  
\begin{eqnarray}
&H_{ph}&=\sum_{i}\sum_{s}\omega_{s}b^{\dag}_{is}b_{is} \label{eq_H_ph}, \\
&H_{el-ph}&=\sum_{i}\sum_{s}\sum_{\ell,\ell'}\sum_{\sigma}g^{\ell\ell'}_{s}
d^{\dag}_{i\ell\sigma}d_{i\ell'\sigma}(b^{\dag}_{is}+b_{is}), \label{eq_H_elph}
\end{eqnarray}
where $b_{is}$ is the annihilation
operator for the phonon of the mode $s$ ($=A_{1g}$, $B_{1g}$,
$E_{g}^1$ and $E_{g}^2$) at the site $i$, $\omega_{s}$ is the phonon frequency  and $g^{\ell\ell'}_{s}$
is the electron-phonon coupling.  We note that $E_{g}^1$ and $E_g^2$
correspond to the oscillation  along the $x$- and $y$-axis in the
$E_g$ mode. As following ref. 68, we expand the
electrostatic potential variance for Fe-3$d$ electrons from the four surrounding
As$^{4-}$ ions due to the oscillations of the Fe atoms for the $B_{1g}$ and
$E_g$ modes in the
displacement of the Fe atoms  up to the first order and  expand that in the $x$,
$y$ and $z$ coordinates up to the second order.  The resulting 
electron-phonon coupling matrix elements of the $B_{1g}$ and $E_g$
phonons are given as follows, 
\begin{eqnarray}
&&\sqrt{3}g^{15}_{E_g^1}=g^{25}_{E_g^1}=g^{34}_{E_g^1} \nonumber \\
&=&-\sqrt{3}g^{14}_{E_g^2}=g^{24}_{E_g^2}=-g^{35}_{E_g^2} \\
&&g^{44}_{B_{1g}}=-g^{55}_{B_{1g}}=\sqrt{3}/2g^{12}_{B_{1g}}, \\
&&g^{\ell\ell'}_{s}=g^{\ell'\ell}_{s}, \\
&&g^{\ell\ell'}_{s}=0 \quad \quad (\mathrm{otherwise}).
\end{eqnarray}
In addition, we also consider the electron-phonon coupling for the
$A_{1g}$ phonon,
\begin{equation}
g^{\ell\ell'}_{A_{1g}} = g^{\ell\ell}_{A_{1g}}\delta_{\ell,\ell'} 
\end{equation}

Within the RPA\cite{takimoto,mochizuki,yada}, 
the spin susceptibility $\hat{\chi}^s(q)$ and 
the charge-orbital susceptibility $\hat{\chi}^c(q)$ 
are given in the $50\times50$
matrix representation as follows\cite{yamakawa_1,yamakawa_2,yamakawa_4,yamakawa_5}, 
\begin{eqnarray}
\hat{\chi}^s(q)&=&[\hat{1}-\hat{\chi}^{(0)}(q)\hat{S}]^{-1}\hat{\chi}^{(0)}(q) \label{eq_chis},\\
\hat{\chi}^c(q)&=&[\hat{1}+\hat{\chi}^{(0)}(q)\hat{C}]^{-1}\hat{\chi}^{(0)}(q) \label{eq_chic}
\end{eqnarray}
with the noninteracting susceptibility
\begin{eqnarray}
&&\chi^{(0)~\alpha,\beta}_{\ell_1\ell_2,\ell_3\ell_4}(q)=
-\frac{T}{N}\sum_{k}G^{\beta\alpha}_{\ell_3\ell_1}(k)G^{\alpha\beta}_{\ell_2\ell_4}(k+q) \label{eq_chi0},
\end{eqnarray}
where  $\alpha$, $\beta$ ($=$$A,B$) represent two
Fe sites, $\ell$ represents Fe 3$d$ orbitals,
$\hat{G}(k)=[(i\varepsilon_n+\mu)\hat{1}-\hat{H}_0(\bm{k})]^{-1}$
is the noninteracting Fe-$3d$ electron Green's function in the
10$\times$10 matrix representation, $\mu$ is the chemical potential, 
$\hat{H}_0(\bm{k})$ is the kinetic part of the Hamiltonian with
 the momentum $\bm{k}$ in eq. (\ref{d-p}),
 $k=(\bm{k},i\varepsilon_n)$, $q=(\bm{q},i\nu_m)$ and
$\varepsilon_n=(2n+1)\pi T$ and $\nu_m=2m\pi T$ are the fermionic and
bosonic Matsubara frequencies, respectively. 
It is noted that when  the largest eigenvalue
$\lambda_{\mathrm{spin}}$ ($\lambda_{\mathrm{c-o}}$) of
 $\hat{\chi}^{(0)}(q)\hat{S}$ $(-\hat{\chi}^{(0)}(q)\hat{C})$ reaches unity,
 the magnetic (charge-orbital) instability occurs.

 In the RPA, generally, we need to  collect all the ring- and
 ladder-type Feynman diagrams, where the
 bare vertices for the spin 
 and charge-orbital 
 susceptibilities $\hat{S}$ and $\hat{C}$ in eqs. (\ref{eq_chis}) and
 (\ref{eq_chic}) are
 given by\cite{kontani} 
\begin{eqnarray} 
(\hat{S})^{\alpha,\beta}_{\ell_1\ell_2,\ell_3\ell_4}&=&(\hat{U}^s)^{\alpha,\beta}_{\ell_1\ell_2,\ell_3\ell_4}, \label{eq_s}\\
(\hat{C})^{\alpha,\beta}_{\ell_1\ell_2,\ell_3\ell_4}&=&(\hat{U}^c)^{\alpha,\beta}_{\ell_1\ell_2,\ell_3\ell_4} \nonumber \\
 &-&2\delta_{\alpha\beta}\sum_{s}g^{\ell_2\ell_1}_{s}g^{\ell_3\ell_4}_{s}D_{s}(i\nu_m) \label{eq_c},  
\end{eqnarray}
where the $D_{s}(i\nu_m)=2\omega_{s}/(\nu_m^2+\omega_{s}^2)$
is the local phonon Green's function for the
mode $s$ and $\hat{U}^s$ and $\hat{U}^c$ are the bare vertices due to
the Coulomb interaction given as follows,
\begin{equation}
\hat{U}^s~(\hat{U}^c)= \left\{
\begin{array}{@{\,} l @{\,} c}
U~(U) & (\alpha=\beta,~\ell_1=\ell_2=\ell_3=\ell_4)\\
U'~(-U'+2J) & (\alpha=\beta,~\ell_1=\ell_3\ne\ell_2=\ell_4)\\
J~(2U'-J) & (\alpha=\beta,~\ell_1=\ell_2\ne\ell_3=\ell_4)\\
J'~(J')& (\alpha=\beta,~\ell_1=\ell_4\ne\ell_2=\ell_3)\\
0 & (\mathrm{otherwise})
\end{array} \right. . \label{eq_U}
\end{equation} 
 In eqs. (\ref{eq_s}) and (\ref{eq_c}), we neglect the ladder terms for the phonon-mediated interaction.
 This is valid  when  the condition $\omega_{s}\ll E_F$ is satisfied, where 
  $E_F$ is the Fermi energy,  because the ladder terms
 are proportional to the 
 power of $\omega_s/ E_F$ in the weak coupling regime under the
 condition  $\omega_s\ll E_F$\cite{migdal}.
The vertex corrections including the ladder
  terms, however, play 
  significant roles in the intermediate and strong coupling regime even
  though $\omega_s\ll E_F$\cite{hague}.  The effect of the ladder
  terms will be discussed in Sec. III D.

 The linearized Eliashberg equation is given
by 
\begin{eqnarray}
\lambda_{\mathrm{sc}}\Delta^{\alpha\beta}_{\ell\ell'}(k)&=&-\frac{T}{N}\sum_{k'}
\sum_{\ell_1\ell_2\ell_3\ell_4}\sum_{\alpha',\beta'}V^{\alpha,\beta}_{\ell\ell_1,\ell_2\ell'}(k-k')\ \ \ \ \ \ \nonumber\\
&\times&G^{\alpha' \alpha}_{\ell_3\ell_1}(-k')
\Delta^{\alpha'\beta'}_{\ell_3\ell_4}(k') G^{\beta' \beta}_{\ell_4\ell_2}(k') \label{eq_gap}, \label{gapeq}
\end{eqnarray}
where $\Delta^{\alpha\beta}_{\ell\ell'}(k)$ is the  gap function
and $V^{\alpha,\beta}_{\ell_1\ell_2,\ell_3\ell_4}(q)$ is the
effective pairing interaction for the spin-singlet state.
Within the RPA\cite{takimoto,mochizuki,yada}, $V^{\alpha,\beta}_{\ell_1\ell_2,\ell_3\ell_4}(q)$ 
is given in the  $50\times50$
matrix,
\begin{equation}
\hat{V}(q)=\frac{3}{2}\hat{S}\hat{\chi}^s(q)\hat{S}-\frac{1}{2}\hat{C}\hat{\chi}^c(q)\hat{C}
+\frac{1}{2}\left(\hat{S}+\hat{C}\right)\label{eq_veff_s}.
\end{equation}
The linearized Eliashberg equation (\ref{eq_gap}) is solved to
obtain the gap function $\Delta^{\alpha\beta}_{\ell\ell'}(k)$
with the eigenvalue $\lambda_{\mathrm{sc}}$. At $T=T_c$, the largest eigenvalue $\lambda_{\mathrm{sc}}$
becomes unity. 
In the present paper, we only focus on the case with $x=0.1$, where
the superconductivity is observed in the 1111 system\cite{kamihara}.
For simplicity, we set $T=0.02\mathrm{eV}$ and 
$\omega_{A_{1g}}=\omega_{B_{1g}}=\omega_{E_{g}^1}=\omega_{E_{g}^2}=\omega_0=0.02\mathrm{eV}$
in 
the present study\cite{boeri,rahlenbeck,zhao}.  As ref. 68, we assume
 $g^{15}_{E_g^1}=g^{34}_{E_g^1}=g^{14}_{E_g^2}=g^{35}_{E_g^2}=g^{12}_{B_{1g}}=0$ and
 also set 
$g^{25}_{E_g^1}=g^{24}_{E_g^2}=g^{44}_{B_{1g}}=-g^{55}_{B_{1g}}=g^{\ell\ell}_{A_{1g}}=g$.

We use $32\times32$ 
$\bm{k}$ point meshes and 512 Matsubara frequencies 
($-511 \pi T\le \varepsilon_n \le 511\pi T$) in the numerical calculations for
eqs. (\ref{eq_chis})-(\ref{eq_veff_s}), and perform the summation of the
momentum and the frequency in eqs. (\ref{eq_chi0}) and (\ref{eq_gap}) by
using the fast Fourier 
transformation (FFT).
 Here and hereafter, we
measure the energy in units of eV.
\begin{figure}[t]
\begin{center}
\includegraphics[width=70.0mm]{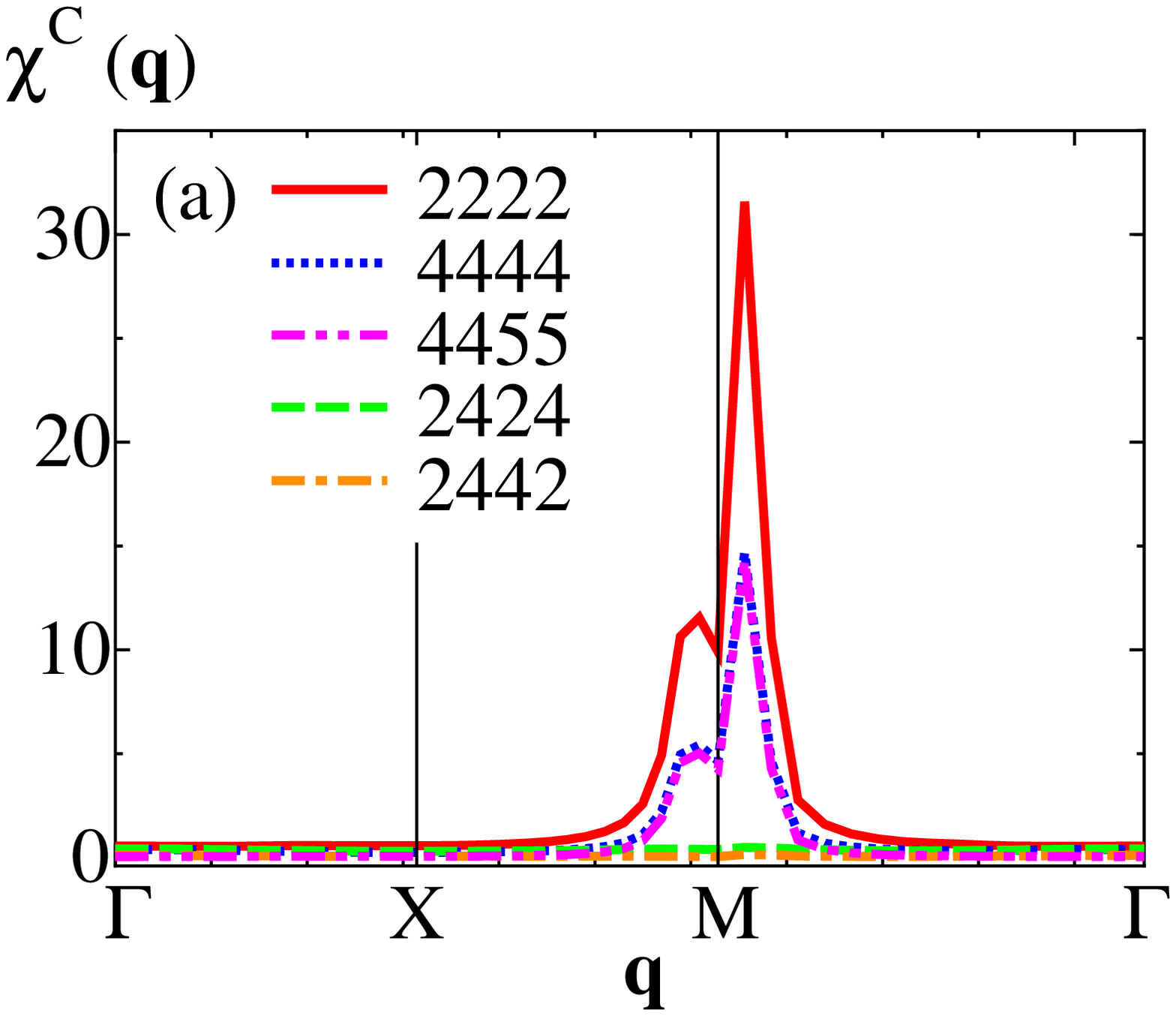}
\includegraphics[width=70.0mm]{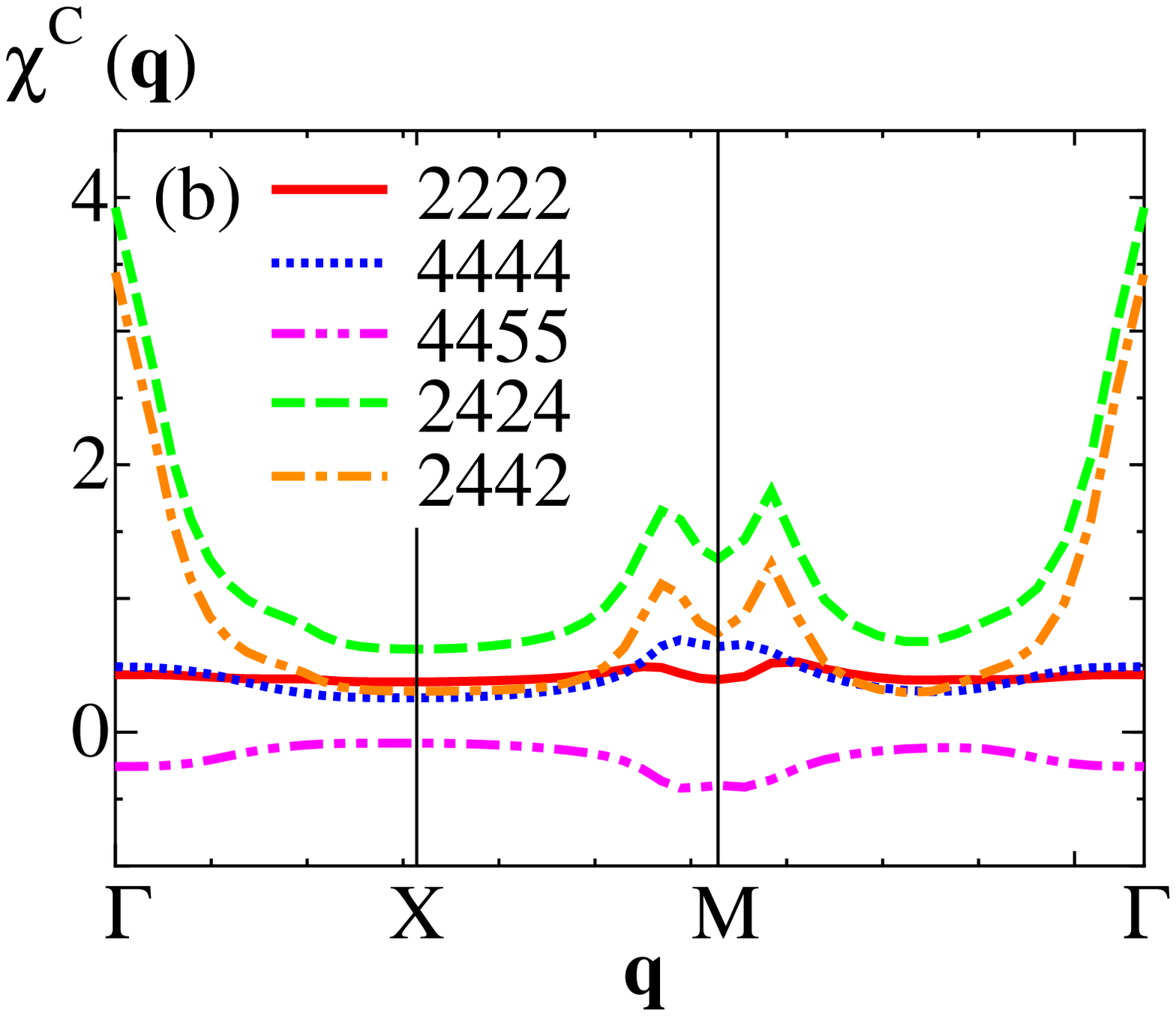}
\caption{(Color online)  Several components of the charge-orbital
 susceptibility $\hat{\chi}^c(\bm{q},0)$ for $U'=0$ and
 $2g^2/\omega_0=0.31$ (a) and  those for $U'=1.0$ and
 $2g^2/\omega_0=0.45$ at $J=J'=0.1$ (b). \label{chi}}
\end{center}
\end{figure}

\section{Calculated Results}
\subsection{Charge-orbital susceptibility}
As presented in eqs. (\ref{eq_s}) and (\ref{eq_c}), where we neglect
the ladder terms, the spin susceptibility is not 
affected by the electron-phonon interaction and is the same as the
 results shown in our previous papers\cite{yamakawa_1,yamakawa_2,yamakawa_4,yamakawa_5}. Therefore, first, we focus
on the effects of the electron-phonon interaction on the charge-orbital
susceptibility. The effects of
the ladder terms on the spin and charge-orbital susceptibilities
will be discussed in Sec. III D.

The several components of the static charge-orbital susceptibility
$\hat{\chi}^c (\bm{q},0)$ for $U'=0$ and $2g^2/\omega_0=0.31$ and those
for $U'=1.0$ and $2g^2/\omega_0=0.45$ at $J=J'=0.1$ are
shown in
Fig. \ref{chi} (a) and (b), respectively, where the parameters are
chosen to satisfy the condition $\lambda_{\mathrm{sc}}\sim 1$. It is 
found that for $U'=0$ and $2g^2/\omega_0=0.31$, the diagonal  
components of  $\hat{\chi}^c(\bm{q},0)$, especially  
$[\hat{\chi}^c(\bm{q},0)]^{A,A}_{22,22}$,  
 are large  and have   
sharp peaks around $\bm{q}\sim (\pi,\pi)$ which originate from the
nesting between the hole FSs and the electron FSs (see Fig. \ref{FS}), while the off-diagonal components are
small. When $U'=0$ and $2g^2/\omega_0=0.31$, $\lambda_{\mathrm{c-o}}\sim 0.99$
and the charge susceptibility  
$\sum_{\ell,\ell',\alpha,\beta}[\hat{\chi}^c(\bm{q},0)]^{\alpha,\beta}_{\ell\ell,\ell'\ell'}$
 becomes almost divergent, where the charge
fluctuations dominate over the orbital fluctuations.  We note that the
 charge-fluctuations are enhanced  due to the
effects of the $A_{1g}$ phonon.
\begin{figure}[t]
\begin{center}
\includegraphics[height=80.0mm,angle=-90]{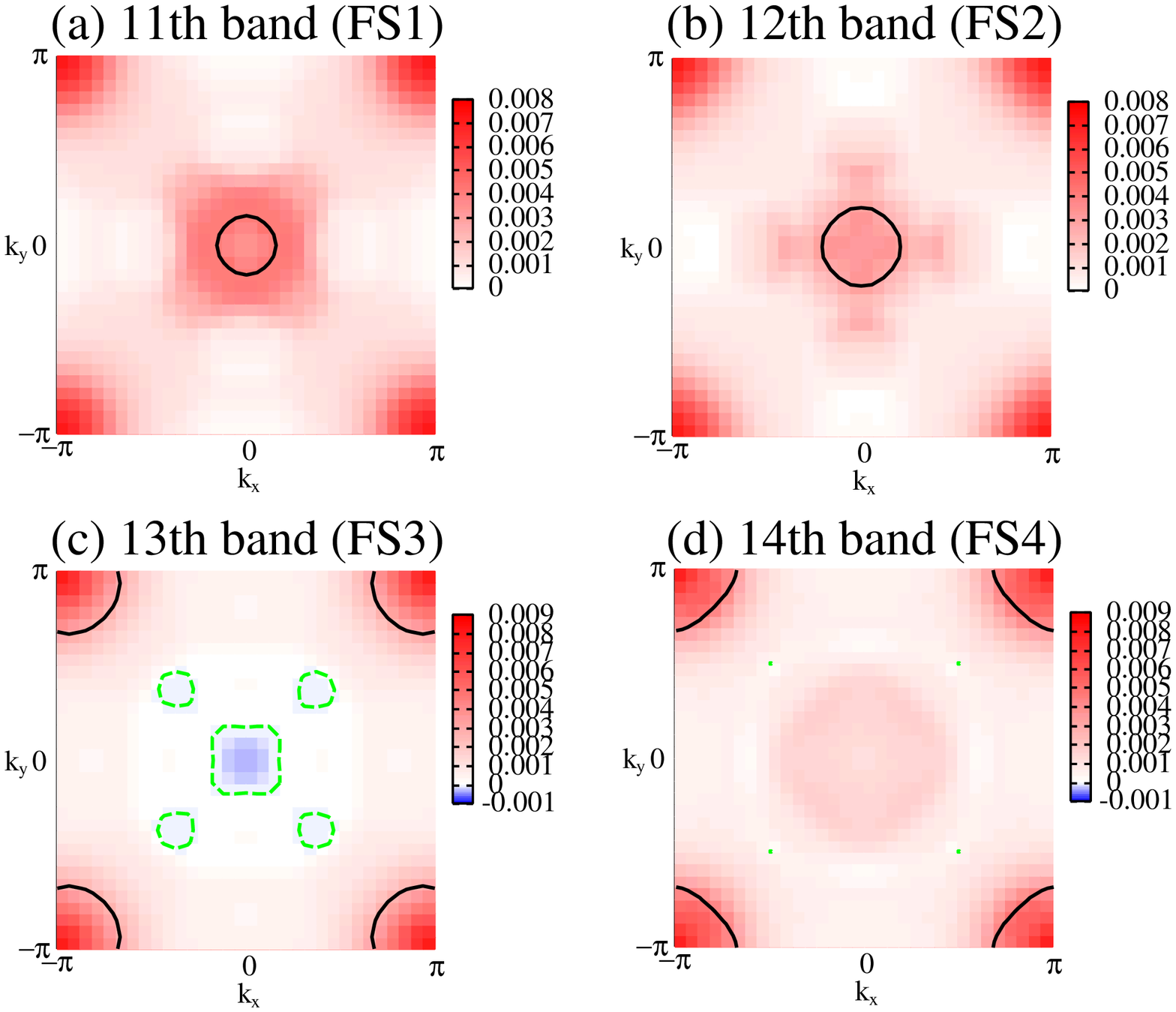}
\caption{(Color online) The diagonal components of the gap function
 $\hat{\Delta} (\bm{k},i\pi T)$ in
 the band representation for $U'=0$, $2g^2/\omega_0=0.31$ and
 $J=J'=0.1$. (a), (b), (c) and (d) correspond to  the 11th,
 12th, 13th and 14th bands, respectively. The solid
 and dashed lines represent the FSs and the nodes of the gap function, respectively. \label{fig_gap_1}}
\end{center}
\end{figure}
\begin{figure}[t]
\begin{center}
\includegraphics[height=80.0mm,angle=-90]{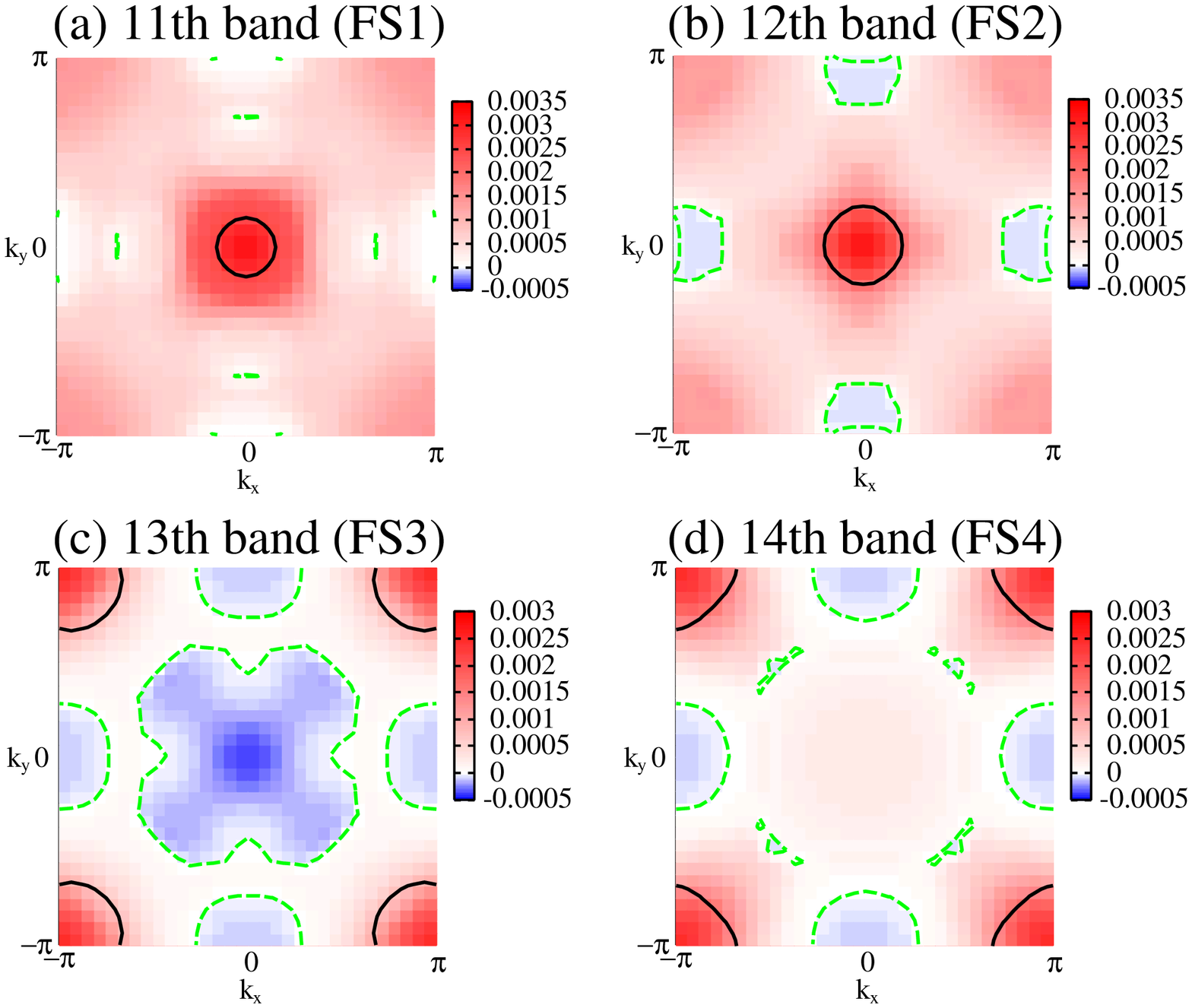}
\caption{(Color online)  The diagonal components of the gap function
 $\hat{\Delta} (\bm{k},i\pi T)$ in
 the band representation for $U'=1.0$, $2g^2/\omega_0=0.45$ and
 $J=J'=0.1$. (a), (b), (c) and (d) correspond to  the 11th,
 12th, 13th and 14th bands, respectively. The solid
 and dashed lines represent the FSs and the nodes of the gap function, respectively. \label{fig_gap_2}}
\end{center}
\end{figure}
\begin{figure}[t]
\begin{center}
\includegraphics[height=80.0mm,angle=-90]{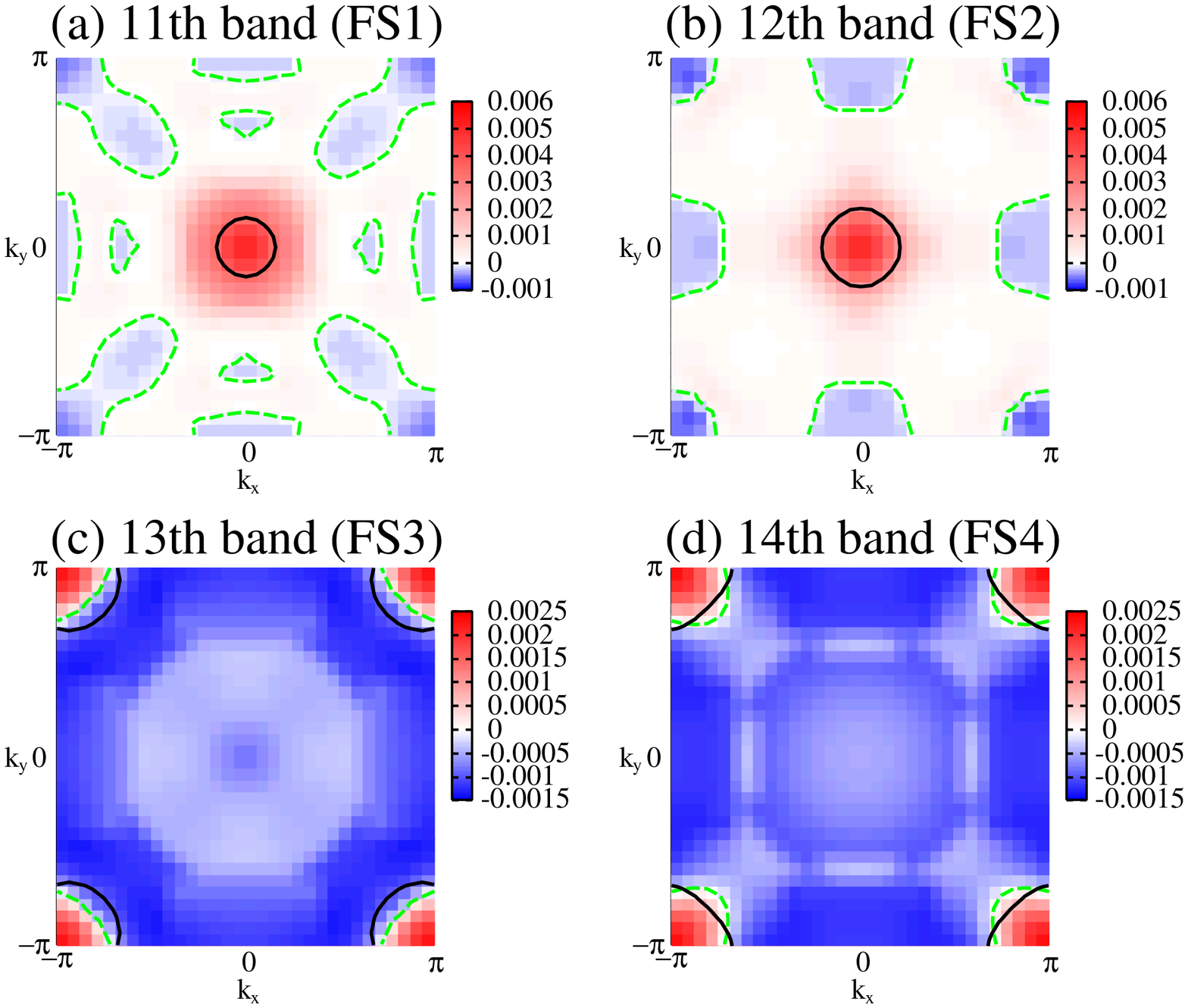}
\caption{(Color online) The diagonal components of the gap function
 $\hat{\Delta} (\bm{k},i\pi T)$ in
 the band representation for $U'=1.5$, $2g^2/\omega_0=0.34$ and
 $J=J'=0.1$. (a), (b), (c) and (d) correspond to the 11th,
 12th, 13th and 14th bands, respectively. The solid
 and dashed lines represent the FSs and the nodes of the gap function, respectively. \label{fig_gap_3}}
\end{center}
\end{figure}

As shown in Fig. \ref{chi} (b),
 for $U'=1.0$ and $2g^2/\omega_0=0.45$, the off-diagonal
components of  $\hat{\chi}^c(\bm{q},0)$, especially
$[\hat{\chi}^c(\bm{q},0)]^{A,A}_{24,24}$ and
$[\hat{\chi}^c(\bm{q},0)]^{A,A}_{24,42}$ , are large and have broad 
 peaks around $\bm{q}=(0,0)$ and $\bm{q}\sim (\pi,\pi)$, 
 while the diagonal components are not so
 large in contrast to the case with $U'=0$ and $2g^2/\omega_0=0.31$ (see
 Fig \ref{chi} (b)). We note that $[\hat{\chi}^c(\bm{q},0)]^{A,A}_{24,24}$ and
 $[\hat{\chi}^c(\bm{q},0)]^{A,A}_{24,42}$  represent the 
transverse orbital fluctuations.  The peaks around $\bm{q}=(0,0)$ and $\bm{q}\sim (\pi,\pi)$ originate from the
 scattering in the electron FSs and the nesting between the hole FSs and
 the electron FSs, respectively. When  $U'=1.0$ and
 $2g^2/\omega_0=0.45$, $\lambda_{\mathrm{c-o}}\sim 0.97$ and  the off-diagonal components
$[\hat{\chi}^c(\bm{q})]^{A,A}_{24,24}$ and
$[\hat{\chi}^c(\bm{q})]^{A,A}_{24,42}$ are largely enhanced, while the charge susceptibility  
$\sum_{\ell,\ell',\alpha,\beta}[\hat{\chi}^c(\bm{q})]^{\alpha,\beta}_{\ell\ell,\ell'\ell'}$
is not 
enhanced because of the  negative contributions of the orbital
susceptibilities such as $[\hat{\chi}^c(\bm{q},0)]^{A,A}_{44,55}$ (see
Fig. \ref{chi}). Then,
 the orbital fluctuations dominate over  the charge
fluctuations in contrast to the
case with $U'=0$ and $2g^2/\omega_0=0.31$.  
We  note that
the off-diagonal components of  the orbital susceptibilities
$[\hat{\chi}^c(\bm{q},0)]^{A,A}_{24,24}$ and 
 $[\hat{\chi}^c(\bm{q},0)]^{A,A}_{24,42}$ are enhanced  due to the
effects of the $E_{g}$ phonon, while $[\hat{\chi}^c(\bm{q},0)]^{A,A}_{44,44}$ and 
 $[\hat{\chi}^c(\bm{q},0)]^{A,A}_{44,55}$ are enhanced  due to the
effects of the $B_{1g}$ phonon. In addition, the orbital fluctuations
are enhanced also due to the effect of the inter orbital Coulomb
interaction $U'$ as shown in our previous paper\cite{yamakawa_5}.

\subsection{Gap function}
By solving the linearized Eliashberg equation, we obtain the gap
function with the lowest matsubara frequency   in the
orbital representation $\Delta^{\alpha\beta}_{\ell\ell'}(\bm{k},i\pi T)$. Then performing the unitary transformation, we
obtain the diagonal components of the gap function
in the band representation $\hat{\Delta}(\bm{k},i\pi T)$. Figs. \ref{fig_gap_1} shows  $\hat{\Delta}(\bm{k},i\pi T)$
for $U'=0$ and
$2g^2/\omega_0=0.31$. It is found that  the pairing symmetry is 
the $s_{++}$-wave
state, where the gap function has no sign change on  the whole FSs. In
this regime,  the  $s_{++}$-wave superconductivity is mediated by 
the charge fluctuations which is enhanced due to the effects of the
$A_{1g}$ phonon as mentioned before (see Fig. \ref{chi} (a)). 

Figs. \ref{fig_gap_2} shows  $\hat{\Delta}(\bm{k},i\pi T)$
for $U'=1.0$ and
$2g^2/\omega_0=0.45$. It is found that  the pairing symmetry is 
the $s_{++}$-wave
state, where the gap function has no sign change on  the whole FSs. In
this regime,  the  $s_{++}$-wave superconductivity is mediated by 
the orbital fluctuations which is enhanced due to the cooperative effects of the
$B_{1g}$, $E_g$ phonons and the inter-orbital Coulomb interaction $U'$
as mentioned before (see Fig. \ref{chi} (b)).  

\begin{figure}[t]
\begin{center}
\includegraphics[width=30mm,angle=-90]{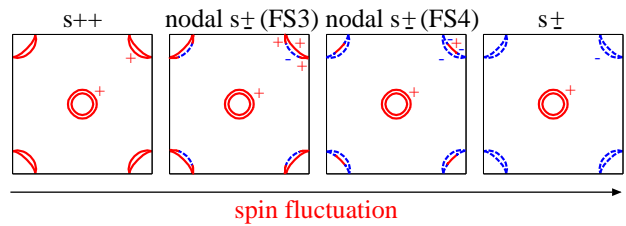}
\caption{(Color online)  The schematic figure of the crossover between the $s_{++}$-wave and the
 nodal $s_{\pm}$-wave states and that between the nodal $s_{\pm}$-wave
 and the
  $s_{\pm}$-wave states. The solid and dashed lines represent the FSs on
 which the signs of the gap functions are positive and negative, respectively.
 \label{fig_cross}}
\end{center}
\end{figure}

In contrast to the above two cases, 
 for $U'=1.5$ and $2g^2/\omega_0=0.34$,  
the gap function have nodes on the FS4 as shown in Fig. \ref{fig_gap_3}. Thus, the pairing
state is the nodal $s_{\pm}$-wave state which  originate from the coexistence of the orbital fluctuations 
with $\bm{q}=(0,0)$ and the spin 
fluctuations with
$\bm{q}\sim(\pi,\pi)$\cite{yamakawa_1,yamakawa_2,yamakawa_4,yamakawa_5}.
 When we further increase the Coulomb interaction, the spin fluctuations dominate
 over the orbital fluctuations resulting in the $s_{\pm}$-wave state, where
 the sign of the gap function between the hole FSs and the electron FSs
 as shown in Fig. \ref{fig_cross}\cite{yamakawa_4,yamakawa_5}. Then, the
 nodal $s_{\pm}$-wave state is observed in the crossover region between the  $s_{\pm}$-wave phase and
 the $s_{++}$-wave phase as shown in Fig. \ref{fig_cross}.

\subsection{Phase diagram}
\begin{figure}[t]
\begin{center}
\includegraphics[width=80mm]{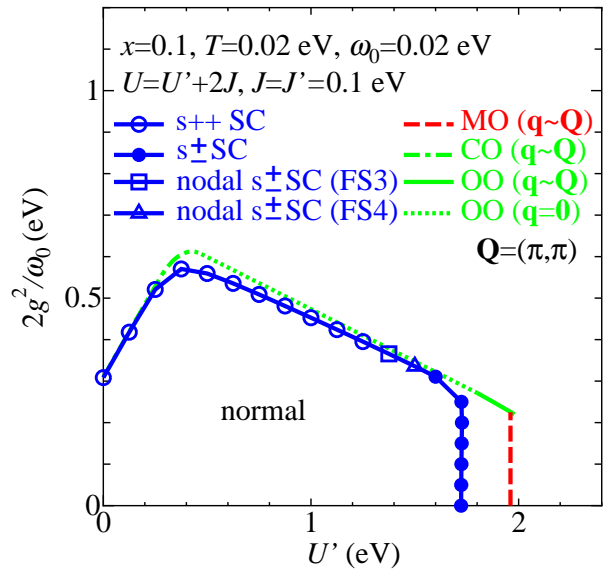}
\caption{(Color online) (a) The phase diagram on the $U'$-$2g^2/\omega_0$ plane for
 $J=J'=0.1$ at
 $x=0.1$, $T=0.02$. The open and solid circles  represent the $s_{++}$-wave and 
 $s_{\pm}$-wave superconducting instabilities, respectively. The open
 triangles and squares represent the  nodal $s_{\pm}$-wave 
 superconducting instabilities whose nodes are on the FS3 and FS4,
 respectively.  The dashed, solid, dotted, and dot-dashed lines show the
 magnetic order with $\bm{q}\sim(\pi,\pi)$, orbital order with
 $\bm{q}\sim(\pi,\pi)$,  orbital order with
 $\bm{q}=(0,0)$, and charge order with $\bm{q}\sim(\pi,\pi)$,
 respectively.
 \label{fig_pd}}
\end{center}
\end{figure}

 The phase diagram on $U'$-$2g^2/\omega_0$ plane is shown in
 Fig. \ref{fig_pd}. It is found that the phase diagram 
 includes the charge, orbital,
 and magnetic order and the superconductivity, where the
 charge-orbital, magnetic and superconducting instabilities 
 are determined by the condition that $\lambda_{\mathrm{c-o}}$,
 $\lambda_{\mathrm{spin}}$ and $\lambda_{\mathrm{sc}}$ reach unity as 
 mentioned in Sec. II, respectively.

 First, we focus on the ordered phase. As shown in Fig. \ref{fig_pd},
 for $0 \le U' < 0.44$, the charge 
 order with $\bm{q}\sim(\pi,\pi)$ takes 
 place at a certain critical value of $2g^2/\omega_0$.  Since both
 the intra- and inter-orbital direct terms $U$  
 and $U'$ suppress the charge fluctuations, the 
 critical values of $2g^2/\omega_0$ increase with  increasing $U'$.
 For
 $0.44 < U' < 1.8$, the ferro orbital order with
 $\bm{q}=(0,0)$ takes place, while the antiferro-like orbital order with  $\bm{q}\sim(\pi,\pi)$ takes
 place for $1.8<U'<1.96$.  Since the inter-orbital direct term $U'$  
  enhances the orbital fluctuations\cite{yamakawa_5}, the 
 critical values of $2g^2/\omega_0$ decrease with  increasing $U'$.  For
   $U'> 1.96$, the stripe-type AFM order with  
$\bm{q}\sim(\pi,\pi)$ takes place  as presented in the previous
 papers\cite{yamakawa_1,yamakawa_2,yamakawa_4,yamakawa_5}. 

 Now, let us bring our attention to the superconductivity.   It is found
 that the pairing 
 symmetry is always 
 $s$-wave and the gap structure sensitively depends on $U'$ and
 $2g^2/\omega_0$.  For $U'\le 1.25$, the
 $s_{++}$-wave superconductivity is realized near the charge and orbital
 ordered phases. The superconductivity near the charge
 ordered phase is mediated by
 the charge fluctuations characterized by $[\hat{\chi}^c
 (\bm{q},0)]^{A,A}_{\ell\ell,\ell\ell}$, while that near the orbital ordered
 phase is 
 mediated by the orbital fluctuations 
 characterized by  $[\hat{\chi}^c
 (\bm{q},0)]^{A,A}_{24,24}$, $[\hat{\chi}^c 
 (\bm{q},0)]^{A,A}_{24,42}$, $[\hat{\chi}^c (\bm{q},0)]^{A,A}_{44,44}$
 and $[\hat{\chi}^c (\bm{q},0)]^{A,A}_{44,55}$ (see Fig. \ref{chi}).  On the other hand, for $U'\ge 1.6$, the
 $s_{\pm}$-wave superconductivity is realized near the orbital and
 magnetic ordered phases, where the spin fluctuations are responsible
 for the superconductivity as presented in the previous
 papers\cite{yamakawa_1,yamakawa_2,yamakawa_4,yamakawa_5}. 
 In addition, the nodal $s_{\pm}$-wave superconductivity is realized between the
 $s_{++}$-wave and $s_{\pm}$-wave superconducting phases.

This crossover
 behavior is naturally explained by the relative strength of the orbital fluctuations and the spin fluctuations. When the spin
 fluctuations are not so strong, the $s_{++}$-wave state is
 realized. With increasing $U'$, the spin fluctuations with
 $\bm{q}\sim (\pi,\pi)$ develop,
 especially $[\hat{\chi}^s (\bm{q},0)]^{A,A}_{22,22}$.  The gap
 function $\Delta^{AA}_{22}(\bm{k},i\pi T)$ has sign change between
 the hole FSs and the electron FSs, where the amplitude of
 $\Delta^{AA}_{22}(\bm{k},i\pi T)$ around the electron FSs is small. On the other hand,  
$\Delta^{AA}_{44}(\bm{k},i\pi T)$ and 
$\Delta^{AA}_{55}(\bm{k},i\pi T)$ have no sign change. As a result, the nodes appear
 on the FS3 which has mainly orbital 2 character.  As the spin
  fluctuations with $\bm{q}\sim(\pi,\pi)$ further increase, the amplitude of
 $\Delta^{AA}_{22}(\bm{k},i\pi T)$ around the electron FSs
 increase and the nodes appear on the FS4 and finally, for $U'\ge 1.6$
 , the fully-gapped $s_{\pm}$-wave state is realized.
 As shown in
 Fig. \ref{fig_cross},  the nodes firstly appear
 on the FS3 and the position of the nodes smoothly moves to FS4 as $U'$
 increases. Finally, the nodes on the FSs disappear and the
 $s_{\pm}$-wave state is realized. 
 
  We note that   the  
 critical value of $2g^2/\omega_0$ at which the orbital order takes
 place decrease with increasing $U'$ as mentioned above, since $U'$ 
 enhances the orbital fluctuations \cite{yamakawa_5}.  Thus, we stress
 that the orbital order and the
 orbital fluctuation-mediated superconductivity are driven by the
 cooperative 
 effects of  the Coulomb interaction and the electron-phonon interaction. 

\begin{figure}[t]
\begin{center}
\includegraphics[width=80mm]{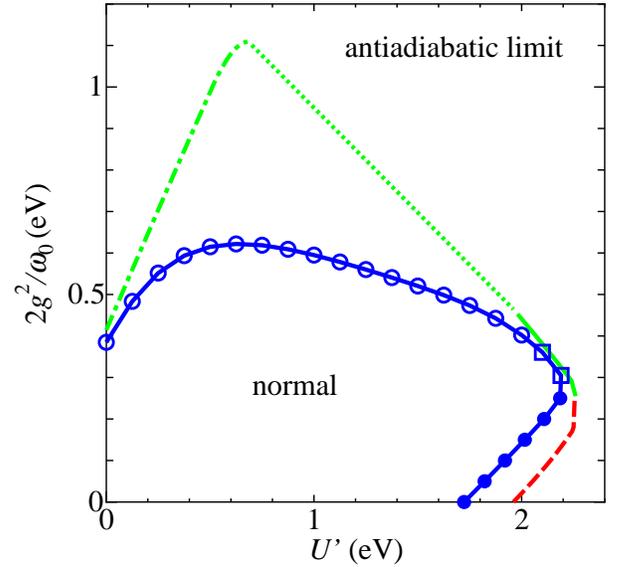}
\caption{(Color online) (a) The phase diagram on the $U'$-$2g^2/\omega_0$ plane for
 $J=J'=0.1$ at
 $x=0.1$, $T=0.02$ in the antiadiabatic
 limit. It is noted that the legends are the same as Fig. \ref{fig_pd} 
 \label{fig_pd_2}}
\end{center}
\end{figure}

\subsection{Antiadiabatic limit}
Finally, we discuss the effects of the ladder type diagrams of the
electron-phonon interaction which are neglected in eqs. (\ref{eq_s}) and
(\ref{eq_c}). 
 In general, it is difficult to include the effects of the ladder
 terms for the phonon-mediated interaction.
Therefore, 
 in this subsection,  
 we study the effects of the ladder  terms by taking the antiadiabatic
 limit ($\omega_{s}\rightarrow \infty$), where the phonon Green's
 function becomes $D_{s}(i\nu_m)\rightarrow 2/{\omega_{s}}$.  The resulting bare
 vertices $\hat{S}$ and $\hat{C}$ are given by, 
\begin{eqnarray} 
(\hat{S})^{\alpha\beta}_{\ell_1\ell_2,\ell_3\ell_4}&=&(\hat{U}^s)^{\alpha\beta}_{\ell_1\ell_2,\ell_3\ell_4} \nonumber \\
&-&2\delta_{\alpha\beta}\sum_{s}g^{\ell_3\ell_1}_{s}g^{\ell_2\ell_4}_{s}/{\omega_s}, \label{eq_aalimit_s}\\
(\hat{C})^{\alpha\beta}_{\ell_1\ell_2,\ell_3\ell_4}&=&(\hat{U}^c)^{\alpha\beta}_{\ell_1\ell_2,\ell_3\ell_4} 
-2\delta_{\alpha\beta}\sum_{s}2 g^{\ell_2\ell_1}_{s}g^{\ell_3\ell_4}_{s}/\omega_{s} \nonumber \\
&+&2\delta_{\alpha\beta}\sum_{s} g^{\ell_3\ell_1}_{s}g^{\ell_2\ell_4}_{s}/{\omega_s} \label{eq_aalimit_c}. 
\end{eqnarray}
Substituting eqs. (\ref{eq_aalimit_s}) and (\ref{eq_aalimit_c}) into
eqs. (\ref{eq_chis}), (\ref{eq_chic}) and (\ref{eq_veff_s}) instead of 
eqs. (\ref{eq_s}) and (\ref{eq_c}), we obtain the
RPA results with including the ladder terms of the electron-phonon interaction. 
Although the antiadiabatic limit is not physically relevant, it is
still a useful point of the reference for getting an overall 
understanding of the physics of the model. We note that, when we apply 
the antiadiabatic limit to the vertex $\hat{C}$ given in eq. (\ref{eq_c}) 
where the ladder terms of the electron-phonon interaction are neglected, 
the phase boundaries for the charge, orbital and magnetic orders shown in 
Fig. \ref{fig_pd} are unchanged  although the superconducting phase boundary
is modified due to the retardation effect of the electron-phonon 
interactions as mentioned below. 

 The phase diagram on $U'$-$2g^2/\omega_0$ plane in the
 antiadiabatic limit is  shown in 
Fig. \ref{fig_pd_2}.  
As shown in  eq. (\ref{eq_aalimit_s}), the spin fluctuation enhanced
due to the 1st term of r.h.s. is suppressed by the 2nd term of
r.h.s.  corresponding to the contribution from the ladder terms 
 of the electron-phonon interaction. 
Similarly, the charge-orbital fluctuation enhanced
due to the 2nd term of r.h.s.  in  eq. (\ref{eq_aalimit_c}) is suppressed by 
the 3rd term of r.h.s. corresponding to the contribution from the ladder terms. 
Therefore, the magnetic, charge and orbital instabilities are 
considered to be suppressed by the ladder terms 
 of the electron-phonon interaction. In fact, as  
shown in Fig. \ref{fig_pd_2}, the critical values of $2g^2/\omega_0$ for
  the charge and orbital orders with including the ladder terms 
are larger than those without the ladder terms shown in 
Fig. \ref{fig_pd}.  
 The critical value of $U'$ for the magnetic  
 order with including the ladder terms is also larger than that without the ladder terms
(see Figs. \ref{fig_pd} and \ref{fig_pd_2}).  

Remarkably, the $s_{++}$-wave superconductivity 
 is observed in a considerably wide parameter region 
 as shown in  Fig. \ref{fig_pd_2}.  
In the antiadiabatic limit, the bare vertex $\hat{C}$ 
given in eq. (\ref{eq_aalimit_c}) 
is independent of the frequency, and then, 
the effective  pairing interaction $\hat{V}(q)$ due to the charge-orbital 
fluctuations becomes attractive for a wide frequency range. 
On the other hand, in the case with the frequency dependent vertex $\hat{C}$
 given in eq. (\ref{eq_c}),  the effective pairing interaction $\hat{V}(q)$  
due to the charge-orbital fluctuations becomes attractive 
only for the low frequency $|\nu_m|<\omega_0$. 
As the result, the $s_{++}$-wave superconductivity due to the 
charge-orbital fluctuations is observed in a relatively narrow region 
as shown in Fig. \ref{fig_pd} for $T=\omega_0=0.02$, 
where $\hat{V}(q)$ is attractive only for $\nu_m$ with $m=0$ 
as $\nu_1=2\pi T >\omega_0$. 
When the temperature is  lowered below $\omega_0$, 
$\hat{V}(q)$ becomes attractive also for $\nu_m$ with $|m|\ge 1$, 
and then, it is expected that the $s_{++}$-wave superconductivity 
 is observed in a wider parameter region also for the 
 case with the frequency dependent vertex.

  \section{Summary and Discussion}
 In summary, we have investigated  the
 two-dimensional 16-band $d$-$p$  model coupled with $A_{1g}$, $B_{1g}$
 and   $E_{g}$ local phonons using the RPA and have obtained the
 phase diagram including the magnetic, charge and orbital ordered phases 
 on the parameter plane of the Coulomb and electron-phonon interactions
 as follows: 
  (1) For weak Coulomb interaction, the charge order with $\bm{q}\sim(\pi,\pi)$ 
      takes place due to the effect of the  electron-phonon interaction 
      with $A_{1g}$ mode. 
  (2) For intermediate Coulomb interaction, 
      the orbital order with  $\bm{q}\sim(0,0)$ takes place 
      due to the cooperative effects of the Coulomb interaction and 
      the electron-phonon interaction with $B_{1g}$ and $E_{g}$ modes. 
      The orbital order with $\bm{q}\sim(\pi,\pi)$ also takes place 
      for relatively larger value of the Coulomb interaction. 
  (3) For strong Coulomb interaction, the stripe-type antiferromagnetic order 
      with $\bm{q}\sim(\pi,\pi)$ takes place due to the effect of the Coulomb 
      interaction. 

 Using the effective pairing interaction obtained from the RPA, 
 we have also solved the linearized Eliashberg equation to obtain 
 the superconducting phase diagram including the three types of 
  $s$-wave pairing as follows: 
  (1) Near the charge ordered phase for weak Coulomb interaction, 
      the $s_{++}$-wave pairing is mediated by the charge fluctuations. 
  (2) Near the orbital ordered phase for intermediate Coulomb interaction, 
      the $s_{++}$-wave pairing is mediated by the orbital fluctuations  
      when the spin fluctuations are not so strong, 
      while the nodal $s_{\pm}$-wave pairing is mediated by both of 
      the orbital and spin fluctuations when the spin fluctuations 
      are rather strong. 
  (3) Near the magnetic ordered phase for strong Coulomb interaction, 
      the $s_{\pm}$-wave pairing is mediated by the spin fluctuations.

Similar phase diagram including the magnetic and orbital ordered phases had 
recently been obtained by Kontani and Onari using the 5-band 
Hubbard-Holstein model\cite{kontani}. They had also discussed the 
superconductivity due to the magnetic and orbital fluctuations 
and have claimed that the $s_{++}$-wave superconductivity is realized 
near the orbital ordered phase, while the $s_{\pm}$-wave superconductivity 
is realized near the magnetic ordered phase, although the detailed 
superconducting phase diagram has not been obtained. 
In the present study, we have explicitly obtained the superconducting phase
 diagram and have found that not only the $s_{++}$-wave but also 
 the nodal $s_{\pm}$-wave superconductivity is realized near the orbital 
 ordered phase in contrast to the prediction in ref. 68. 
In addition, the charge order and the charge fluctuation-mediated 
$s_{++}$-wave superconductivity have been found to take place due to 
the effect of the $A_{1g}$ local phonon which was not been 
considered in ref. 68. 
In early theoretical studies for the copper oxide superconductors, 
the effect of the Coulomb interaction between the $d$ and $p$ 
electrons $U_{pd}$ was found to enhance the charge fluctuations 
which induce the $s$-wave superconductivity\cite{littlewood_2,hirashima}. 
We have also discussed the effect of $U_{pd}$ on the present 
 $d$-$p$ model for the iron-based superconductors and have found that 
 $U_{pd}$ enhances the charge fluctuation-mediated $s_{++}$-wave 
superconductivity. The detailed results will be published in a 
subsequent paper.

It seems that 
both the $s_{\pm}$-wave and the $s_{++}$-wave states
with full superconducting gaps are consistent with various
experiments in the iron-based superconductors as mentioned in Sec. I, 
although the sign of the gap function has not been 
directly observed. However, the recent theoretical studies of 
the nonmagnetic impurity effects\cite{onari} revealed that 
the Anderson's theorem is violated for the $s_{\pm}$-wave superconductivity
in contrast to the experimental results of very weak
$T_c$ suppression in Fe site substitution\cite{kawabata_1,sefat} and
neutron irradiation\cite{karkin}. 
As the impurity potential due to  the Fe-site substitution is considered to be 
 diagonal and local in the orbital basis according to the first  principle
 calculation\cite{kemper}, it is expected that the 
 $s_{++}$-wave state is  more robust against the nonmagnetic impurity 
 than the $s_{\pm}$-wave state. 
In the recent ultrasonic measurements\cite{fernandes,yoshizawa}, 
a remarkable softening of the elastic constant is observed at 
low temperature down to $T_c$ 
and is well accounted for by Jahn-Teller modes coupled with 
strong orbital fluctuations\cite{yoshizawa}. 
In addition, the weak $T$-dependence of $1/T_1T$\cite{nakano,nakai_2} 
above $T_c$ in the electron-doped compounds is considered to indicate the weak spin fluctuations. 
Thus, the $s_{++}$-wave state due to the orbital fluctuations 
 seems to be responsible for the fully gapped superconductivity 
 in the iron-based superconductors.

In BaFe$_2$(As$_{1-x}$P$_x$)$_2$,  
the recent 
 field-angle resolved specific heat\cite{matsuda} and the ARPES 
 measurements\cite{shimojima} suggest that the superconducting 
 gap function has vertical line nodes along  
 the $k_z$-axis on  the electron FSs. 
 This nodal superconductivity seems to correspond to the nodal 
 $s_{\pm}$-wave state obtained in the present study.  
   In the previous works\cite{kuroki_3,maier,kariyado}, 
  the similar nodal $s_{\pm}$-wave states 
  have been obtained in the crossover region between 
  the $s_{\pm}$-wave phase and the $d$-wave phase 
  when the different modes of the spin fluctuations coexist. 
 This is a striking contrast to the case with the present study 
 where the nodal $s_{\pm}$-wave state is realized 
 in the crossover region between the $s_{++}$-wave phase and the
 $s_{\pm}$-wave phase 
 when the strong orbital and spin fluctuations coexist. 
 If the fully gapped superconductivity widely observed in the iron-based 
 superconductors is the $s_{++}$-wave state, 
 it is natural to consider that the nodal superconductivity observed 
 in BaFe$_2$(As$_{1-x}$P$_x$)$_2$ is  the nodal $s_{\pm}$-wave state 
 obtained in the present study.


   In the present and previous papers\cite{yamakawa_5}, 
   we have shown that 
   the electron-phonon interaction plays important roles for the 
   iron-based superconductors in cooperation with the Coulomb interaction. 
   Actually, the Raman spectroscopies indicate the large electron-phonon 
   interaction\cite{rahlenbeck}. The 
  large value of the Gr\"{u}neisen parameter\cite{budko} and the drastic 
  softening of the elastic constant\cite{fernandes,yoshizawa} observed  in
  BaFe$_{2-x}$Co$_x$As$_2$ also indicate the large electron-lattice coupling. 
  Remarkably, the recent ultrasonic measurements revealed that 
  the softening of the elastic  constant $C_{44}$ is much larger than 
  $(C_{11}-C_{12})/2$ and continues down to $T_c$\cite{yoshizawa}, 
  where the temperature dependence of  the elastic  constant is well 
  accounted for by Jahn-Teller modes which couple with 
  the orbital fluctuation between $d_{yz}$ and $d_{zx}$ orbitals: 
  $[\hat{\chi}^c(q)]^{\alpha,\beta}_{44,44}
  -[\hat{\chi}^c(q)]^{\alpha,\beta}_{44,55}$.  
  Since this type of the orbital fluctuation is enhanced due to the 
  electron-phonon interaction with $B_{1g}$ mode, 
  we may expect that the effects of the $B_{1g}$ phonon is most dominant 
  for the elastic softening and the superconductivity. 
  In fact, the first principle calculation\cite{boeri} and the Raman 
  spectroscopies\cite{rahlenbeck,zhao} suggest that the frequency of the $B_{1g}$ 
  phonon is lower than the $A_{1g}$ and $E_g$ phonons, although 
  the same frequencies are assumed in the present study for simplicity. 
  Therefore, we need further investigation of the electron-phonon interaction 
  with including the more realistic effects  such as the mode dependence
  of the  phonon frequencies and the coupling constants and the phonon dispersions 
  which have not been considered in this paper. 
   


\begin{acknowledgments}
The authors thank M. Yoshizawa, 
M. Sato, H. Kontani, S. Onari, T. Nomura, H. Ikeda, K. Kuroki and Y. Yanase for useful
comments and discussions.      This work was partially supported by the
Grant-in-Aid for Scientific Research from the Ministry of Education,
Culture, Sports, Science and Technology.  The authors Y. Yanagi and
 Y. Yamakawa are supported by the Grant-in-Aid for JSPS Fellows. 
\end{acknowledgments}

\bibliography{prb8}
\end{document}